\documentclass[10pt,letterpaper]{article}
\usepackage{opex3}

\begin{document}

\title{Fiber-based cryogenic and time-resolved spectroscopy of PbS quantum dots}

\author{Matthew T. Rakher$^1$, Ranojoy Bose$^{2,3}$, Chee Wei Wong$^2$, and Kartik
Srinivasan$^1$}

\address{$^1$Center for Nanoscale Science and Technology, National Institute of Standards and Technology, Gaithersburg, MD 20899-6203 \\ $^2$Optical Nanostructures Laboratory, Center for
Integrated Science and Engineering, Solid-State Science and
Engineering and Mechanical Engineering, Columbia University, New
York, NY 10027\\
$^3$Current Address: Institute for Research in Electronics and
Applied Physics, University of Maryland, College Park, MD 20742}

\email{matthew.rakher@gmail.com} 



\begin{abstract}
PbS quantum dots are promising active emitters for use with
high-quality Si nanophotonic devices in the telecommunications-band.
Measurements of low quantum dot densities are limited both because
of low fluorescence levels and the challenges of single photon
detection at these wavelengths.  Here, we report on methods using a
fiber taper waveguide to efficiently extract PbS quantum dot
photoluminescence.  Temperature dependent ensemble measurements
reveal an increase in emitted photons concomitant with an increase
in excited-state lifetime from 58.9 ns at 293 K to 657 ns at 40 K.
Measurements are also performed on quantum dots on high-$Q$
($>10^5$) microdisks using cavity-resonant, pulsed excitation.
\end{abstract}

\ocis{(300.6280) Spectroscopy, fluorescence and luminescence;
(230.5590) Quantum-well, -wire and -dot devices; (250.5230)
Photoluminescence
} 




\section{Introduction}

Lead-salt colloidal quantum dots (QDs) \cite{ref:Wise_ACR00} such as
PbS QDs are interesting active emitters due to the fact that they
can be integrated with high quality, Si-based nanophotonic devices
designed to operate near 1550 nm
\cite{ref:Takahashi_OE_09,ref:Srinivasan7}. These devices can
potentially be used for applications such as ultra-low threshold
lasers \cite{ref:Strauf2} as well as fundamental studies of
light-matter interaction \cite{ref:Raimond}. While significant
strides along these lines have been made with self-assembled InAs
QDs embedded in GaAs-based devices
\cite{ref:Srinivasan16,ref:Rakher_PRL_09}, the optical quality
factors ($Q$s) obtained are usually an order of magnitude lower
\cite{ref:Michael} than what can be achieved in Si, SiO$_2$, or SiN
\cite{ref:Takahashi_OE_09}. In addition, the fabrication of
electronic and micromechanical systems in Si-based materials is much
more mature than in GaAs.  Thus, the introduction of an active
emitter with Si-based optical microcavities and waveguides is a very
active field of research
\cite{ref:Polman_JAP97,ref:Fang_Bowers_OE05}. In contrast to InAs
quantum dots, PbS QDs have longer radiative lifetimes by 2 to 3
orders of magnitude \cite{ref:SargentAM05} and reduced quantum
efficiencies by 1 to 2 orders of magnitude when dried
\cite{ref:SteckelAM03}, yielding a photon emission rate that is as
much as five orders of magnitude worse. Combined with the challenges
of single-photon detection in the near-infrared
\cite{ref:Hadfield_nphoton}, performing spectroscopic measurements
of low densities of PbS QDs can be difficult and photon collection
efficiencies must be as high as possible.

Previous work
\cite{ref:Fushman_APL05,ref:WuAPL07,ref:PattantyusNANO09,ref:Bose_OE_09}
with lead-salt QDs on microcavities has focused on using free-space
optics and microscope objectives to collect emission, but this is
limited by the planar geometry of the microcavities. Recently, it
was shown that coupling to photonic crystal cavities through a fiber
taper waveguide is an efficient method to extract emitted photons
\cite{ref:Rakher_APL_2010}. In addition, it was demonstrated that
the QDs did not degrade the $Q$ of the cavity up to
$Q\approx3\times10^4$.  Here, we present further spectroscopic
techniques for PbS QDs based upon efficient collection using a fiber
taper waveguide (FTW). In particular, we perform cryogenic
photoluminescence (PL) studies of low-densities of QDs dried
directly onto an FTW, enabling efficient measurement in a range of
environmental conditions.  Also, we demonstrate PL and time-resolved
PL measurements of QDs dried onto Si microdisk cavities and show
that $Q$s up to $10^5$ are unaffected by the QDs. As discussed in
the last section of this letter, single PbS QD measurements, like
those performed on InAs QDs, will be quite challenging due to the
aforementioned detection difficulties and use of the techniques
described here will be necessary to make such experiments more
feasible.

\section{Experimental Setup}
\begin{figure}[t]
\centerline{\includegraphics[width=11cm,clip=true]{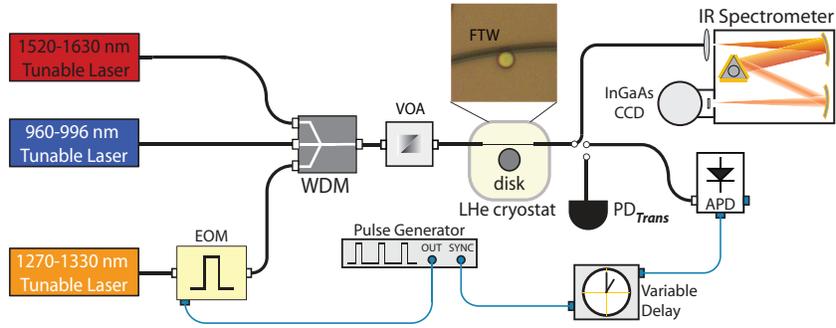}}
\caption{Experimental setup composed of tunable-wavelength laser
sources (980 nm, 1310 nm, and 1550 nm), electro-optic intensity
modulator (EOM), wavelength division multiplexer (WDM), variable
optical attenuator (VOA), fiber taper waveguide (FTW), cryostat,
timing electronics, and measurement devices (spectrometer,
photodiode (PD), and InGaAs/InGaP single photon counting avalanche
photodiode (APD)).} \label{fig:setup}
\end{figure}

The PbS QDs used here are chemically synthesized and stored in a
solution of chloroform \cite{Evident}.  Per the manufacturer's
specification in solution, these QDs are 5.3 nm $\pm$ 0.5~nm in
diameter and have strong optical absorption from 400 nm to 1400 nm.
The peak in the emission spectrum in solution is expected near 1450
nm with a width of 150 nm.  Experiments are performed using a fiber
taper waveguide \cite{ref:Srinivasan7}, which is a single mode
optical fiber whose diameter has been adiabatically reduced from 125
$\mu$m to $\approx1$~$\mu$m over a length of 10 mm by heating with a
hydrogen torch. Typical losses due to fabrication are less than 0.5
dB. Light propagating through the FTW is used to evanescently excite
photoluminescence (PL) from PbS QDs or to probe cavity resonances in
Si microdisk structures. Because the FTW begins and terminates as
single mode fiber, a variety of tunable laser sources can be easily
introduced with a wavelength division multiplexer into the setup as
shown in Fig.~\ref{fig:setup}. Each source is required for a
different measurement.  The 960 nm to 996 nm laser is used to
generate excitons far above the photoluminescence at 1500 nm,
enabling measurement of the complete emission spectrum.  The 1270 nm
to 1330 nm laser is used for both CW excitation and pulsed
excitation for time-resolved measurements. The 1520 nm to 1630 nm
laser is used to measure microdisk cavity resonances in transmission
near the emission band of the QDs.

After variable optical attenuation, the sources are introduced
through fiber feedthroughs into one of two setups. The first is a
liquid He flow cryostat for measurements of PbS QDs on FTWs as a
function of temperature, while the second is a N$_2$-rich room
temperature enclosure for measurements of PbS QDs dried onto optical
cavities. Temperature measurements are performed in the cryostat
using a silicon diode attached adjacent to the copper mount holding
the FTW.
The FTW is controlled within the room temperature enclosure with
high resolution (50 nm minimum step size) stages to enable accurate
positioning with respect to microdisk cavities. For transmission
measurements, the signal is measured directly with an InGaAs
photodiode. For PL measurements, the signal is measured using a
monochromator with a liquid-nitrogen-cooled InGaAs array or an
InGaAs/InGaP single photon counting avalanche photodiode (APD). The
pulsed source near 1300~nm required for time-resolved measurements
is created using an electro-optic intensity modulator driven by a
pulse generator. These instruments combine to create
wavelength-tunable (1300 nm to 1320 nm) 22.8 fJ, 2.1 ns pulses at a
user-defined repetition rate of 1 KHz to 1 MHz with an on/off
extinction ratio of 28.1 dB.  It is important to note that a
standard Ti:sapphire pulsed laser would be inappropriate for the
measurement because the typical repetition rate (80 MHz) is too fast
compared to the expected decay time of QDs (100 ns to 1 $\mu$s) and
the EOM-based pulsed excitation offers substantially greater
flexibility in repetition rate, wavelength, and pulse length.  Using
an faster pulse generator it would be possible to create pulses as
short as 100 ps, making future Purcell effect measurements possible.
Because the APD must run in a gated-detection mode, the detection
gate is synchronized to the arrival of the optical pulse with a
delay generator triggered by the pulse generator as shown in
Fig.~\ref{fig:setup}.  Time-resolved measurements are accomplished
by changing the delay of the APD gating with respect to the arrival
of the PL pulse and integrating the counts for a specified time.  If
the expected decay times were shorter than the width of the APD gate
(100 ns maximum), the measurement could be performed in a
multi-channel way using a standard time-correlated counting board.

\section{Cryogenic and Time-resolved Measurements of PbS QDs dried on FTW}
\begin{figure}[t]
\centerline{\includegraphics[width=11cm,clip=true]{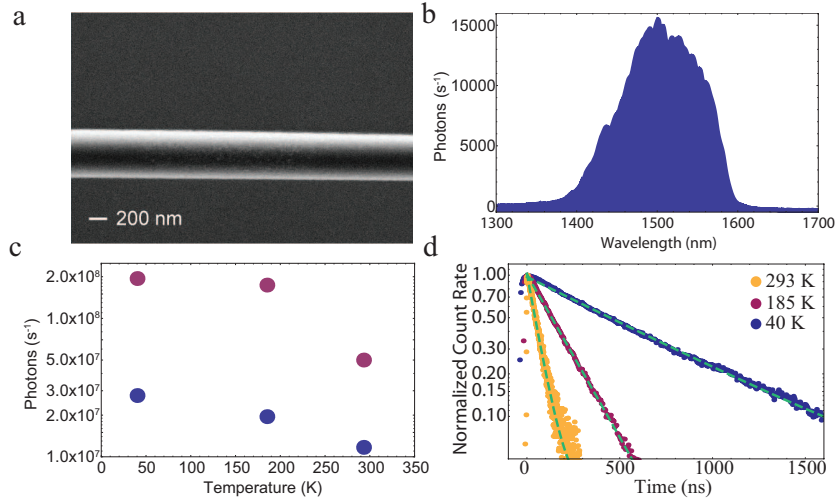}}
\caption{(a) Scanning electron microscope image of 450 nm diameter
fiber taper waveguide with PbS QDs dried on surface (not visible).
(b) Spectrum (60 s integration time) of QD PL at 40~K under 6.7 mW
of 980 nm excitation. (c) Comparison of total PL counts
($\lambda>1400$~nm) under 1 $\mu$W (blue) and 10 $\mu$W (maroon)
1310~nm excitation at 40~K, 185~K, and 293~K. Errors are contained
within the point size.  (d) Time-resolved PL traces taken at 40~K,
185~K, and 293~K with 2.1~ns, 22.8 fJ pulses at 1310~nm with 190 kHz
repetition rate. The extracted decay times (errors given by 95~$\%$
confidence interval) are 58.9~ns$~\pm3.7$~ns, 189.1~ns$~\pm4.4$~ns,
and 657~ns$~\pm10$~ns for 293~K, 185~K, and 40~K respectively.}
\label{fig:taper}
\end{figure}

In the first set of measurements, PbS QDs in chloroform solution are
diluted to a concentration of 0.5 mg/mL.  Then, $\approx$10 $\mu$L
drops are dried on a glass slide overlayed with the FTW, resulting
in a low density ($\approx$ 1000~$\mu$m$^{-2}$ as measured by
scanning electron microscopy) drying onto the FTW. Previous
theoretical \cite{ref:LeKien.pra.72.032509,ref:Davanco} and
experimental work \cite{ref:Nayak,ref:Gregor_OE_09,
ref:Vetsch_prl_10} has shown that FTWs can be efficient channels for
the collection of spontaneous emission from nearby, optically-active
sources such as atoms, ions, or QDs. In fact, it was shown
theoretically \cite{ref:LeKien.pra.72.032509} that an
optimally-sized FTW of diameter $\approx \lambda/4$ could collect
approximately 28~$\%$ of the total light from a nearby emitter. In
the experiments here, the FTW is used both for efficient evanescent
excitation and collection of subsequent PL. The FTW is tapered to a
nearly-optimal diameter of $\approx$450 nm, see
Fig.~\ref{fig:taper}a, to optimize collection efficiency near 1500
nm while minimizing scattering loss. For the FTW shown in
Fig.~\ref{fig:taper}a, the loss induced by the tapering was
$\approx$5~dB, much more than for a diameter of 1~$\mu$m (0.5 dB),
and is due to a breakdown of adiabaticity at these small length
scales. The FTW with PbS QDs is loaded into the He-flow cryostat for
PL measurements at 293 K, 185 K, and 40 K. PL spectroscopy is
performed by exciting the PbS QDs at 980~nm or 1310~nm and measuring
the emitted photons with the InGaAs array after spectral dispersion
by a monochromator as shown in Fig.~\ref{fig:setup}.
A complete 40~K PL spectrum excited by 6.7 mW at 980 nm is shown in
Fig.~\ref{fig:taper}b and shows the center of the emission near
1500~nm with a width $\approx175$~nm. For comparison at different
temperatures, the total collected PL with $\lambda>1400$~nm excited
by 1~$\mu$W (blue) and 10~$\mu$W (maroon) at 1310 nm is shown in
Fig.~\ref{fig:taper}c for temperatures of 40~K, 185~K, and 293~K.
The data clearly show increasing PL count rates for decreasing
temperatures.  Along with the increase in count rates, the center
wavelength was measured to shift slightly to longer wavelengths with
decreasing temperature, an effect that has been previously measured
for the PbS QDs \cite{ref:Turyanska_APL_07}, and serves to verify
the temperature assignments.

To further investigate the dynamics of photoluminescence in these
QDs at different temperatures, APDs used in conjunction with 190 kHz
pulsed excitation at 1310~nm (see Fig.~\ref{fig:setup}) enabled
measurement of the excited state lifetime as shown for each
temperature in Fig.~\ref{fig:taper}d, where photon counts for
$\lambda>1400$~nm, are displayed as a function of the delay between
the arrival of the optical pulse and the gating of the APD. Each
point on the curve was sequentially obtained after 2 s of
integration time, followed by a 2 s dark count only measurement for
background subtraction. As shown clearly in Fig.~\ref{fig:taper}d,
the exponential decay of the excited population changes drastically
with temperature resulting in extracted lifetimes (errors given by
95~$\%$ confidence interval) of 58.9~ns$~\pm3.7$~ns,
189.1~ns$~\pm4.4$~ns, and 657~ns$~\pm10$~ns for 293~K, 185~K, and
40~K respectively.  The simultaneous increase in the measured count
rates with an increase in lifetime for decreasing temperatures is
strongly indicative of a significant non-radiative decay channel
that decreases with temperature.  Notably, the data does not fit
well to the temperature dependence of multi-phonon relaxation
\cite{ref:Layne_PRB_77} arising from coupling to
longitudinal-optical (LO) phonons in PbS at $E_{LO}= 26$~meV
\cite{ref:Turyanska_APL_07}, which indicates some other form of
energy transfer. Such energy transfer effects are common phenomena
in other solid-state optical emitters and somewhat limits their
utility at room-temperature. Specifically, fundamental light-matter
studies need to be performed at temperatures with the best possible
quantum coherence.

\section{Time-resolved Measurements of PbS QDs on Si Microdisks}
\begin{figure}[t]
\centerline{\includegraphics[width=11cm,clip=true]{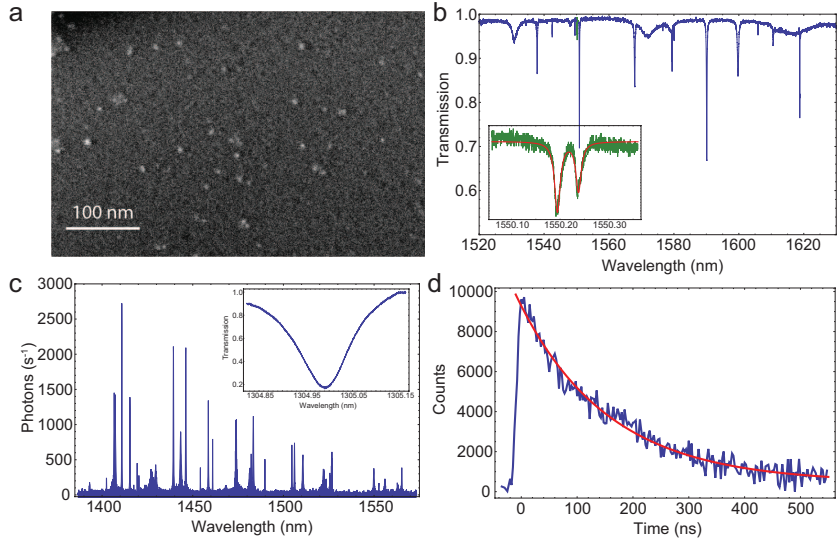}}
\caption{(a) SEM image of PbS QDs deposited on the surface of a Si
microdisk, corresponding to an areal density of $\approx2000$
$\mu$m$^{-2}$. (b) Transmission measurement of the FTW coupled to a
Si microdisk from 1520~nm to 1630~nm, showing several cavity modes.
A $Q\approx 1.1\times10^5$ doublet is shown in green and enlarged
with fit in inset. (c) PL measurement of QD emission collected by
the FTW in contact with the microdisk under pulsed excitation at
1304 nm. Inset: Transmission spectrum of the pump mode at 1304 nm
with $Q\approx 9 \times 10^3$ and $\Delta T = 0.92\pm0.02$.  (d)
Time-resolved PL decay trace measured with the InGaAs/InGaP APD.}
\label{fig:disk}
\end{figure}

To investigate coupling to cavities, a low density of PbS QDs were
spun directly onto a wafer containing 4.5~$\mu$m diameter Si
microdisk cavities. These cavities were fabricated using standard
silicon-on-insulator processing with a 250 nm thick device layer. An
SEM of the surface of one such microdisk is shown in
Fig.~\ref{fig:disk}a after PbS QD spin, resulting in a QD density of
approximately 2000~$\mu$m$^{-2}$. In order to determine the optical
quality factors, the FTW was positioned in the near-field of the
cavities to enable efficient, resonant interaction with light
propagating down the fiber.  As in Fig.~\ref{fig:setup}, a tunable
1500~nm band laser was swept in wavelength, coupled to the microdisk
via the FTW, and measured with a low-noise InGaAs photodiode.  The
resulting transmission as a function of wavelength from 1520~nm to
1630~nm is shown in Fig.~\ref{fig:disk}b, where several sharp modes
are clearly visible. For this particular cavity, $Q$s ranged from
$5\times10^3$ to $1.1\times10^5$ (in green near 1550 nm).  The best
measured $Q$s (device not shown) were $2\times10^5$ and were likely
limited by sidewall roughness induced in the etching procedure.
Errors in fitted $Q$ values arise from fitting the data and are
smaller than the last given digit.  Similar to
\cite{ref:Rakher_APL_2010}, microdisks were measured before and
after QD deposition to determine if there was any effect on the $Q$
factor.  Up to $Q\approx 2\times10^5$, there was no measurable $Q$
degradation caused by the presence of the QDs.

Because of the rich mode structure of the disk and the optical
properties of Si, high-$Q$ modes can be measured into the absorption
band of the QDs and down to 1200 nm.  One such mode is shown in
transmission in the inset of Fig.\ref{fig:disk}c with $Q\approx
9\times10^3$.  The FTW enables efficient coupling to this mode with
coupling depth $\Delta T=0.92\pm0.02$, just short of critical
coupling, ensuring almost complete transfer of power.  By pumping on
this mode, only QDs on the circumference of the disk are excited by
the circulating pump.  The 1300 nm tunable laser combined with the
EOM enable cavity-resonant pulsed excitation.  Using 2.1 ns pulses
at 1304 nm (resonant with the mode) with 1 MHz repetition rate, the
QDs emit a PL signal under 79 nW of average power as shown in the
spectrum in Fig.~\ref{fig:disk}c. The significant reduction in
average power required to obtain a bright PL spectrum is due to the
buildup of power in the cavity mode. In the spectrum, the modes of
the microdisk in the emission band are clearly visible and dress the
broad PL signal of the QDs. Incorporation of the gated APD and
requisite timing electronics enables a time-resolved measurement of
the PL decay as shown in Fig.~\ref{fig:disk}d.  The lifetime
extracted from this room-temperature measurement is 160~ns $\pm$
20~ns. This lifetime is longer than that measured for the QDs dried
directly on the FTW at room-temperature, but is of the same order of
magnitude.  In general for these QDs, we measure markedly different
excited state lifetimes for different samples. This implies strong
density-dependent or environmentally-dependent decay dynamics such
as F\"{o}rster resonant energy transfer (FRET) \cite{ref:BoseNL08}.
These effects need to be investigated further and will be the focus
of future work.

\section{Discussion and Conclusion}
We have demonstrated techniques for efficient collection of photons
emitted by PbS QDs in the 1.5~$\mu$m band using FTWs in different
environments as well as on microdisk cavities.  Using these
collection strategies, it is an important exercise to consider if it
is experimentally feasible to detect the single photon emission from
a single PbS QD in a lifetime measurement and in a Hanbury-Brown and
Twiss setup, much like what is done with InAs QDs. In the collection
geometries presented here, the fraction of spontaneous emission
collected in the measurement mode should be on the order of 1~$\%$
to 28~$\%$ \cite{ref:LeKien.pra.72.032509}, neglecting any
substantial Purcell enhancement. Coupled with assumed radiative
lifetimes as long as 700~ns, photoluminescence photon rates from a
single PbS QD should reach as high as $\approx 2\times10^5$ s$^{-1}$
under saturated continuous wave excitation if the radiative
efficiency at low temperature approaches unity.  However, these
seemingly high count rates are mitigated by the difficulties of
single photon detection in the near-infrared
\cite{ref:Hadfield_nphoton}.  In particular, for the InGaAs APD used
in our experiments, the optimum detection parameters for a
single-channel lifetime measurement like that found in
Fig.~\ref{fig:taper}d and Fig.~\ref{fig:disk}d are a 20~$\%$
detection efficiency, 100~ns gate width, and 10~$\mu$s dead time for
a 1 MHz trigger rate. These settings have a dark count rate of
$\approx1.7 \times 10^3$~s$^{-1}$, yielding a signal to noise ratio
of $\approx$35.4 Hz$^{-1/2}$. Experimentally, this means that the
excited-state decay will be observed with a dynamic range of 35.4 if
each temporal point is integrated for 1~s, corresponding to a
measurable decay over a time period of $\approx3.5$ times the decay
constant. Compared to an upconversion, multi-channel Si APD
measurement for InAs QDs, the dynamic range is approximately a
factor of 30 times worse \cite{ref:Rakher_NPhot_2010}.

More important than a lifetime measurement is that of the second
order intensity correlation $g^{2}(\tau)$, where
\begin{equation}
g^{(2)}(\tau) = \frac{\langle
a^\dag(t)a^\dag(t+\tau)a(t+\tau)a(t)\rangle}{\langle a^\dag(t+\tau)
a(t+\tau) \rangle \langle a^\dag(t) a(t) \rangle}
\end{equation}
and $a$ ($a^{\dag}$) is the photon annihilation (creation) operator,
which yields information about the non-classicality of the emitted
photon stream. Specifically, a measured value of $g^{(2)}(0)<0.5$
proves that the field is dominantly composed of single photons.
Using a standard Hanbury-Brown and Twiss setup, a minimum measurable
$g^{(2)}(0)$ value for a PbS QD would be $\approx$0.60 using a
10~$\%$ detection efficiency, 2.5~ns gate width, and 10~$\mu$s dead
time for a 1 MHz trigger rate. While this is a reasonably low value
(proving the field is non-classical but not single photon), the
signal to noise under these measurement conditions is only $\approx
0.011$~Hz$^{-1/2}$, requiring more than two hours of integration
time to achieve unity signal to noise.  A signal to noise of
$\approx 0.16$~Hz$^{-1/2}$ could be obtained by increasing the gate
width to 50~ns, but the minimum measurable $g^{(2)}(0)$ under these
conditions is $\approx 0.81$. In summary, under the best possible
collection conditions and assuming perfect radiative efficiency, it
is not possible to measure $g^{(2)}(0)<0.5$ using commercially
available detectors.  In addition, the assumption of perfect
radiative efficiency does not match experimental observation, even
at cryogenic temperatures. This could be caused by several factors
including blinking \cite{ref:Rakher_APL_2010,ref:ChungPRB04} and
poor radiative efficiency out of solution \cite{ref:SteckelAM03}.
Therefore, it appears that single QD measurements at these
wavelengths may remain elusive until higher optical quality QDs can
be regularly fabricated \cite{ref:Pietryga}. However, once these QDs
are realized, measurement of single photons will likely require
advanced detector technologies such as frequency upconversion
\cite{ref:Rakher_NPhot_2010} or superconducting single photon
detectors \cite{ref:Stevens_APL06}. Nonetheless, the efficient and
flexible measurement techniques presented here will be of great use
towards the development of PbS QDs as active emitters coupled to
high-quality nanophotonic devices in the telecommunications-band.

The authors acknowledge fabrication support from D.~L. Kwong and M.
Yu at the Institute of Microelectronics in Singapore, funding
support from NSF ECCS 0747787, the Nanoscale Science and Engineering
Initiative under NSF Award Number CHE-0641523, and the New York
State Office of Science, Technology, and Innovation.

\begin{thebibliography}{10}
\newcommand{\enquote}[1]{``#1''}
\expandafter\ifx\csname url\endcsname\relax
  \def\url#1{{#1}}\fi
\expandafter\ifx\csname urlprefix\endcsname\relax\def\urlprefix{}\fi

\bibitem{ref:Wise_ACR00}
F.~Wise, \enquote{Lead salt quantum dots: The limit of strong
quantum
  confinement,} Acc. Chem. Res. {\bf 33}, 773--780 (2000).

\bibitem{ref:Takahashi_OE_09}
Y.~Takahashi, Y.~Tanaka, H.~Hagino, T.~Sugiya, Y.~Sato, T.~Asano,
and S.~Noda,
  \enquote{Design and demonstration of high-Q photonic heterostructure
  nanocavities suitable for integration,} Opt. Express {\bf 17},
  18\,093--18\,102 (2009).

\bibitem{ref:Srinivasan7}
K.~Srinivasan, P.~E. Barclay, M.~Borselli, and O.~Painter,
  \enquote{{Optical-fiber-based measurement of an ultrasmall volume, high-$Q$
  photonic crystal microcavity},} Phys. Rev. B {\bf 70}, 081\,306R (2004).

\bibitem{ref:Strauf2}
S.~Strauf, K.~Hennessy, M.~T. Rakher, Y.-S. Choi, A.~Badolato, L.~C.
Andreani,
  E.~L. Hu, P.~M. Petroff, and D.~Bouwmeester, \enquote{{Self-tuned quantum dot
  gain in photonic crystal lasers},} Phys. Rev. Lett. {\bf 96}, 127\,404
  (2006).

\bibitem{ref:Raimond}
J.~Raimond, M.~Brune, and S.~Haroche, \enquote{{Manipulating quantum
  entanglement with atoms and photons in a cavity},} Rev. Mod. Phys. {\bf 73},
  565--582 (2001).

\bibitem{ref:Srinivasan16}
K.~Srinivasan and O.~Painter, \enquote{{Linear and nonlinear optical
  spectroscopy of a strongly coupled microdisk-quantum dot system},} Nature
  (London) {\bf 450}, 862--865 (2007).

\bibitem{ref:Rakher_PRL_09}
M.~T. Rakher, N.~G. Stoltz, L.~A. Coldren, P.~M. Petroff, and
D.~Bouwmeester,
  \enquote{Externally Mode-Matched Cavity Quantum Electrodynamics with
  Charge-Tunable Quantum Dots,} Phys. Rev. Lett. {\bf 102}, 097\,403 (2009).

\bibitem{ref:Michael}
C.~Michael, K.~Srinivasan, T.~Johnson, O.~Painter, K.~Lee,
K.~Hennessy, H.~Kim,
  and E.~Hu, \enquote{{Wavelength- and material-dependent absorption in GaAs
  and AlGaAs microcavities},} Appl. Phys. Lett. {\bf 90}, 051\,108 (2007).

\bibitem{ref:Polman_JAP97}
A.~Polman, \enquote{Erbium implanted thin film photonic materials,}
J. Appl.
  Phys. {\bf 82}, 1--39 (1997).

\bibitem{ref:Fang_Bowers_OE05}
H.~Park, A.~Fang, S.~Kodama, and J.~Bowers, \enquote{Hybrid silicon
evanescent
  laser fabricated with a silicon waveguide and III-V offset quantum well,}
  Opt. Express {\bf 13}, 9460--9464 (2005).

\bibitem{ref:SargentAM05}
E.~H. Sargent, \enquote{Infrared Quantum Dots,} Advanced Materials
(Weinheim,
  Ger.) {\bf 17}, 515 (2004).

\bibitem{ref:SteckelAM03}
J.~S. Steckel, S.~Coe-Sullivan, V.~Bulovi\'{c}, and M.~G. Bawendi,
\enquote{1.3
  $\mu$m to 1.55 $\mu$m Tunable Electroluminescence from PbSe Quantum Dots
  Embedded within an Organic Device,} Advanced Materials (Weinheim, Ger.) {\bf
  15}, 1862 (2003).

\bibitem{ref:Hadfield_nphoton}
R.~H. Hadfield, \enquote{{Single-photon detectors for optical
quantum
  information applications},} Nature Photonics {\bf {3}}, {696--705} ({2009}).

\bibitem{ref:Fushman_APL05}
I.~Fushman, D.~Englund, and J.~Vu\v{c}kovi\'{c}, \enquote{Coupling
of PbS
  quantum dots to photonic crystal cavities at room temperature,} Appl. Phys.
  Lett. {\bf 87}, 241\,102 (2005).

\bibitem{ref:WuAPL07}
Z.~Wu, Z.~Mi, P.~Bhattacharya, T.~Zhu, and J.~Xu, \enquote{Enhanced
spontaneous
  emission at 1.55 mu m from colloidal PbSe quantum dots in a Si photonic
  crystal microcavity,} Applied Physics Letters {\bf 90}, 171\,105 (2007).

\bibitem{ref:PattantyusNANO09}
A.~G. Pattantyus-Abraham, H.~Qiao, J.~Shan, K.~A. Abel, T.-S. Wang,
F.~C. J.~M.
  van Veggel, and J.~F. Young, \enquote{Site-Selective Optical Coupling of PbSe
  Nanocrystals to Si-Based Photonic Crystal Microcavities,} Nano Letters {\bf
  9}, 2849 (2009).

\bibitem{ref:Bose_OE_09}
R.~Bose, J.~Gao, J.~F. McMillan, A.~D. Williams, and C.~W. Wong,
  \enquote{Cryogenic spectroscopy of ultra-low density colloidal lead
  chalcogenide quantum dots on chip-scale optical cavities towards single
  quantum dot near-infrared cavity QED,} Opt. Express {\bf 17},
  22\,474--22\,483 (2009).

\bibitem{ref:Rakher_APL_2010}
M.~T. Rakher, R.~Bose, C.~W. Wong, and K.~Srinivasan,
\enquote{Spectroscopy of
  1.55 $\mu$m PbS quantum dots on Si photonic crystal cavities with a fiber
  taper waveguide,} Applied Physics Letters {\bf 96}, 161\,108 (2010).

\bibitem{Evident}
Purchased from Evident Technologies and identified in this paper to
foster
  understanding, without implying recommendation or endorsement by NIST.

\bibitem{ref:LeKien.pra.72.032509}
F.~Le~Kien, S.~Dutta~Gupta, V.~I. Balykin, and K.~Hakuta,
\enquote{Spontaneous
  emission of a cesium atom near a nanofiber: Efficient coupling of light to
  guided modes,} Phys. Rev. A {\bf 72}, 032\,509 (2005).

\bibitem{ref:Davanco}
M.~{Davan\c co} and K.~Srinivasan, \enquote{Efficient spectroscopy
of single
  embedded emitters using optical fiber taper waveguides,} Opt. Express {\bf
  17}, 10\,542--10\,563 (2009).

\bibitem{ref:Nayak}
K.~Nayak, P.~Melentiev, M.~Morinaga, F.~Kien, V.~Balykin, and
K.~Hakuta,
  \enquote{{Optical nanofiber as an efficient tool for manipulating and probing
  atomic fluorescence},} Opt. Express {\bf 15}, 5431--5438 (2007).

\bibitem{ref:Gregor_OE_09}
M.~Gregor, A.~Kuhlicke, and O.~Benson, \enquote{Soft-landing and
optical
  characterization of a preselected single fluorescent particle on a tapered
  optical fiber,} Opt. Express {\bf 17}, 24\,234--24\,243 (2009).

\bibitem{ref:Vetsch_prl_10}
E.~Vetsch, D.~Reitz, G.~Sagu\'e, R.~Schmidt, S.~T. Dawkins, and
  A.~Rauschenbeutel, \enquote{Optical Interface Created by Laser-Cooled Atoms
  Trapped in the Evanescent Field Surrounding an Optical Nanofiber,} Phys. Rev.
  Lett. {\bf 104}, 203\,603 (2010).

\bibitem{ref:Turyanska_APL_07}
L.~Turyanska, A.~Patan\`{e}, M.~Henini, B.~Hennequin, and N.~R.
Thomas,
  \enquote{Temperature dependence of the photoluminescence emission from
  thiol-capped PbS quantum dots,} Appl. Phys. Lett. {\bf 90}, 101\,913 (2007).

\bibitem{ref:Layne_PRB_77}
C.~B. Layne, W.~H. Lowdermilk, and M.~J. Weber, \enquote{Multiphonon
relaxation
  of rare-earth ions in oxide glasses,} Phys. Rev. B {\bf 16}, 10--20 (1977).

\bibitem{ref:BoseNL08}
R.~Bose, J.~F. McMillan, J.~Gao, K.~M. Rickey, C.~J. Chen, D.~V.
Talapin, C.~B.
  Murray, and C.~W. Wong, \enquote{Temperature-Tuning of Near-Infrared
  Monodisperse Quantum Dot Solids at 1.5 $\mu$m for Controllable F\"{o}rster
  Energy Transfer,} Nano Letters {\bf 8}, 2006 (2008).

\bibitem{ref:Rakher_NPhot_2010}
M.~T. {Rakher}, L.~{Ma}, O.~{Slattery}, X.~{Tang}, and
K.~{Srinivasan},
  \enquote{{Quantum transduction of telecommunications-band single photons from
  a quantum dot by frequency upconversion},} Nature Photonics {\bf 4}, 786--791
  (2010).

\bibitem{ref:ChungPRB04}
I.~Chung and M.~G. Bawendi, \enquote{Relationship between single
quantum-dot
  intermittency and fluorescence intensity decays from collections of dots,}
  Phys. Rev. B {\bf 70}, 165\,304 (2004).

\bibitem{ref:Pietryga}
J.~M. Pietryga, D.~J. Werder, D.~J. Williams, J.~L. Casson, R.~D.
Schaller,
  V.~I. Klimov, and J.~A. Hollingsworth, \enquote{Utilizing the Lability of
  Lead Selenide to Produce Heterostructured Nanocrystals with Bright, Stable
  Infrared Emission,} J. Am. Chem. Soc. {\bf 130}, 4879--4885 (2008).

\bibitem{ref:Stevens_APL06}
M.~J. Stevens, R.~H. Hadfield, R.~E. Schwall, S.~W. Nam, R.~P.
Mirin, and J.~A.
  Gupta, \enquote{Fast lifetime measurements of infrared emitters using a
  low-jitter superconducting single-photon detector,} Appl. Phys. Lett. {\bf
  89}, 031\,109 (2006).

\end{thebibliography}
\end{document}